\documentclass[a4paper,11pt]{article}
\usepackage{jheppub} 
\usepackage{lineno}
\usepackage{comment}
\usepackage{subfiles}
\usepackage{mathtools}
\usepackage{siunitx}
\usepackage[normalem]{ulem}
\usepackage{graphicx}
\usepackage{subcaption}
\usepackage{float}

\arxivnumber{1234.56789} 

\title{Demonstrating topological identification capabilities of the NEXT experiment at low pressure}

\author[12]{J.~Waiton,}
\author[1]{H.~Almaz\'an,}
\author[12]{B.~Palmeiro,}
\author[6]{G.~Mart\'inez-Lema,}
\author[12]{R.~Guenette,}
\author[2]{V.~\'Alvarez,}
\author[3]{L.~Arazi,}
\author[4]{I.J.~Arnquist,}
\author[5]{F.~Auria-Luna,}
\author[6]{S.~Ayet,}
% ...
\affiliation[1]{
Donostia International Physics Center, BERC Basque Excellence Research Centre, Manuel de Lardizabal 4, San Sebasti\'an / Donostia, E-20018, Spain}
\affiliation[2]{
Instituto de Instrumentaci\'on para Imagen Molecular (I3M), Centro Mixto CSIC - Universitat Polit\`ecnica de Val\`encia, Camino de Vera s/n, Valencia, E-46022, Spain}
\affiliation[3]{
Unit of Nuclear Engineering, Faculty of Engineering Sciences, Ben-Gurion University of the Negev, P.O.B. 653, Beer-Sheva, 8410501, Israel}
\affiliation[4]{
Pacific Northwest National Laboratory (PNNL), Richland, WA 99352, USA}
\affiliation[5]{
Department of Organic Chemistry I, Universidad del Pais Vasco (UPV/EHU), Centro de Innovaci\'on en Qu\'imica Avanzada (ORFEO-CINQA), San Sebasti\'an / Donostia, E-20018, Spain}
\affiliation[6]{
Instituto de F\'isica Corpuscular (IFIC), CSIC \& Universitat de Val\`encia, Calle Catedr\'atico Jos\'e Beltr\'an, 2, Paterna, E-46980, Spain}
% ...
\affiliation[12]{
Department of Physics and Astronomy, University of Manchester, Manchester. M13 9PL, United Kingdom}
% ...
\emailAdd{john.waiton@manchester.ac.uk}

\abstract{The NEXT-100 detector is a high-pressure xenon time projection chamber utilising electroluminescence amplification for sub-1\% FWHM energy resolution and topological discrimination, two key attributes required to achieve the NEXT programme's overarching goal of detecting neutrinoless double beta decay ($0\nu \beta \beta$). The detector has completed its first physics run at the Laboratorio Subterr\'aneo de Canfranc (LSC), with xenon at a pressure of $\sim$\qty{4}{bar}.

In this paper we report on the first validation of NEXT-100's topological performance and present the first topological analysis conducted at low pressure within the detector programme. A Monte Carlo study characterising the effects of pressure on track topology is presented, with qualitative agreement observed in data.
We then demonstrate the topological discrimination capabilities for $0\nu \beta \beta$-like events from $^{208}$Tl decays using a cut-based method considered the `baseline' in NEXT's topological programme, upon which all future analyses will improve.
The analysis described applies a set of selection cuts before implementing a background discrimination algorithm that yields a reported signal efficiency for double-electron tracks and background acceptance for single-electron tracks of $75.6\pm 1.9\,\text{(stat.)}\,^{+3.4}_{-4.1}\,\text{(syst.)}\,\text{\%}$ and $14.7\pm 0.4\,\text{(stat.)}\,^{+0.7}_{-0.8}\,\text{(syst.)}\,\text{\%}$. These results demonstrate the excellent topological discrimination capabilities of the NEXT-100 detector in line with NEXT-White, which achieved a signal efficiency and background acceptance of $71.6 \pm 1.5\,\text{(stat.)} \pm 0.3\,\text{(syst.)}\, \text{\%}$ and $20.6 \pm 0.4\,\text{(stat.)} \pm 0.3\,\text{(syst.)}\, \text{\%}$ respectively.}

\begin{document}

\maketitle
\flushbottom
\section{Introduction}

Neutrinoless double beta decay ($0\nu \beta \beta$) is a hypothetical nuclear decay of significant interest in the particle physics community, as its detection would demonstrate the Majorana nature of the neutrino. The decay results in two electrons simultaneously emitted with no outgoing neutrinos (violating lepton number), a final state only plausible if the neutrino is its own antiparticle \cite{furry_0vbb}.

The current best $0\nu \beta \beta$ half-life sensitivity limit is $T_{1/2} > 3.8 \times 10^{26}$yr for $^{136}$Xe \cite{KamLAND_sensitivity}. Further improvements are expected through three main avenues: the increase in target exposure, reduction of background, and improvements in energy resolution.

The NEXT (Neutrino Experiment with a Xenon TPC) programme focuses on the research and development of high-pressure gaseous xenon time projection chambers (TPCs), which achieve sub-\qty{1}{\%} FWHM energy resolution and effective background suppression using topological discrimination \cite{nygren_xetpc, N100_sensitivity}. 

The previous stage in the NEXT programme was NEXT-White \cite{NEW_detector}, a radiopure demonstrator detector hosted at the Laboratorio Subterr\'aneo de Canfranc (LSC), which deployed $\sim$\qty{4.3}{kg} of $^{136}$Xe in the active volume at \qty{10}{bar} and established the technology's capabilities through its stability and geometrical energy corrections \cite{twonu_NEW, Kr_NEW}. 
This enabled a measurement of sub-\qty{1}{\%} FWHM energy resolution near the $Q_{\beta \beta}$ value \cite{HE_NEW}. 
NEXT-White also demonstrated the unique tracking capabilities of the detector technology \cite{topology_NEW}, motivating further research into improving topological discrimination \cite{deconv_NEW, CNN_NEW}.
Together, these strengths enabled a measurement of the $2\nu \beta \beta$ half-life and a limit on the $0 \nu \beta \beta$ half-life \cite{twonu_NEW,zeronu_NEW}. 

The current focus of the programme is the NEXT-100 detector, which aims to set competitive limits for the $0 \nu \beta \beta$ half-life within $^{136}$Xe \cite{N100_sensitivity}, demonstrating the scalability and progress in background reduction over NEXT-White necessary for future tonne-scale detectors \cite{N100_detector}. 
Following a successful commissioning process \cite{N100_EL, N100_HV}, the detector completed a low-pressure (LPR) physics run at $\sim$\qty{4}{bar} (\qty{20}{kg} of xenon depleted of $^{136}$Xe in the active volume). 
Operation at high pressure (HPR), \qty{10}{bar}, is planned to begin shortly.
Following the standard calibration procedure, a low-energy calibration utilising $^{83m}$Kr was completed, demonstrating the overall stability of the detector and the application of geometrical corrections \cite{N100_kr}.
Subsequently, a high-energy calibration utilising the decay spectra of $^{228}$Th was completed, with a sub-\qty{1}{\%} FWHM energy resolution measured in LPR conditions \cite{N100_HE_eres}. 

This paper studies the performance of NEXT-100 in two overlapping contexts. Firstly, it considers the consequences of low pressure for topological reconstruction, a regime that offers unique opportunities and challenges.
Dedicated Monte Carlo (MC) studies are used to understand the differences between LPR and HPR in the context of topology, but current MC simulations do not account for all LPR and detector effects, precluding a direct quantitative comparison to data in this analysis. 

Secondly, the paper presents initial data-driven results from applying a cut-based topological method to NEXT-100 LPR data. This method is based on the equivalent NEXT-White analysis \cite{topology_NEW}, adapted for low pressure, and serves as the topological discrimination (background rejection) baseline for the NEXT-100 detector.
More sophisticated approaches for topological reconstruction have been demonstrated for NEXT-White, such as track deconvolution via the Richardson-Lucy algorithm \cite{deconv_NEW} and improved classification through convolutional neural networks \cite{CNN_NEW}. Efforts to implement these techniques are focused on future HPR runs, where they are more applicable to the nominal NEXT physics programme, primarily the search for $0 \nu \beta \beta$ decay.

The structure of this paper is as follows. Section \ref{sec:N100} provides a simple overview of the NEXT-100 detector and explains the processing methods used to record and digitise charged particle trajectories.
Section \ref{sec:data_and_topo} presents the data runs of interest and the topological algorithms used, providing context for the analysis described in Section \ref{sec:DEP_ana}.
The analysis section is split into two subsections, the first being a study into the effect of pressure with respect to topology, utilising an MC study and data-driven observations.
The second describes steps taken to obtain a clean dataset for study, and the metrics used to quantify the topological signal efficiency and background rejection available within the NEXT-100 detector in the context of $0\nu \beta \beta$-like events. A cut-based topological discrimination method is applied using these metrics, and a parameter scan is implemented to optimise discrimination power.  Section \ref{sec:discussion} discusses the analysis results in the context of prior detectors within the NEXT programme, outlining the benefits and drawbacks of working with differing pressures, and discusses the future outlook of the NEXT programme and gaseous TPCs more broadly. 

\section{NEXT-100}\label{sec:N100}
\subsection{Apparatus}

The NEXT-100 detector is a cylindrical high-pressure gaseous xenon time projection chamber that utilises asymmetric readout technologies; one plane of 3584 silicon photomultipliers (SiPMs, Hamamatsu S13372-1350TE) is used for track reconstruction, and an opposing plane of 60 photomultiplier tubes (PMTs, Hamamatsu R11410-10) is used for energy reconstruction and timestamping, allowing for longitudinal track positioning.
These two planes are considered the tracking and energy planes (TP and EP) respectively. 
The TP SiPMs are distributed across the TP with a pitch of \qty{15.55}{mm} and size of \qty{1.3}{mm}, in comparison to \qty{10}{mm} pitch and \qty{1.0}{mm} size for its predecessor NEXT-White. The 60 EP PMTs (up from 12 for NEXT-White) are ordered hexagonally in a honeycomb structure.  These two planes collect the light produced within the TPC. The TPC volume is divided into three sections as shown in Figure \ref{fig:detector}:

\begin{figure}[htbp]
\centering
\includegraphics[width=\textwidth]{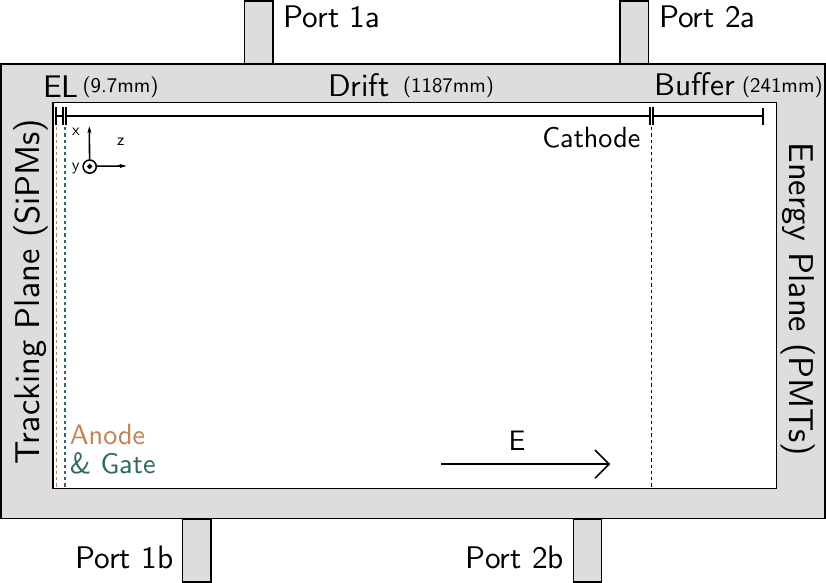}
\caption{A simplified diagram of the NEXT-100 detector from above, visualising the differing regions of the field cage; electroluminescence (EL), drift, and buffer. The calibration ports, cathode, anode and gate are also shown, with the direction of the electric field also provided.}\label{fig:detector}
\end{figure}

\begin{itemize}
    \item the buffer region, spanning \qty{241}{mm} between the EP and cathode, required to separate the large electric potentials at the cathode from the PMTs, 
    \item the drift region, spanning \qty{1187}{mm} between the cathode and the gate, within which a consistent electric field is applied ($\sim$\qty{120}{V/cm}), 
    \item and the electroluminescence (EL) region, spanning \qty{9.7}{mm} between the gate and the anode, where a significantly increased electric field is applied ($\sim$\qty{9000}{V/cm}) to produce electroluminescence light.
\end{itemize}
The cathode, anode and gate consist of photoetched hexagonal meshes, each with an optical transparency of 90\% \cite{N100_EL}. Each aforementioned region has a common inner diameter of \qty{983}{mm}, defined by the lateral surface, which consists of polytetrafluoroethylene (PTFE) plates coated in tetra-phenylbutadiene (TPB), which acts as a wavelength shifter to improve light collection efficiency \cite{TPB_VUV, PTFE_NEXT}. A more comprehensive description of the detector is found in \cite{N100_detector}.

\subsection{Light detection \& processing}\label{ssec:light_detec_proc}
The detection and reconstruction of charged particle trajectories within the detector rely on two distinct light emissions. The first is the \textit{S1} signal generated by the initial scintillation light from charged particles.
This light is wavelength-shifted using TPB from \qty{175}{nm} (vacuum ultraviolet, or VUV) to $\sim$\qty{430}{nm}  and detected by the PMTs as the `prompt' signal within nanoseconds of the scintillation, marking the start of an event \cite{xenon_vuv, TPB_VUV}.

Simultaneously, the charged particles ionise electrons along their path, which are carried by a uniform electric field towards the EL region. There, they are accelerated, producing VUV photons via electroluminescence amplification \cite{electroluminescence}. These VUV photons are wavelength-shifted and detected on both the tracking and energy planes as the \textit{S2} signal.

\begin{figure}[ht]
\centering
\includegraphics[width=\textwidth]{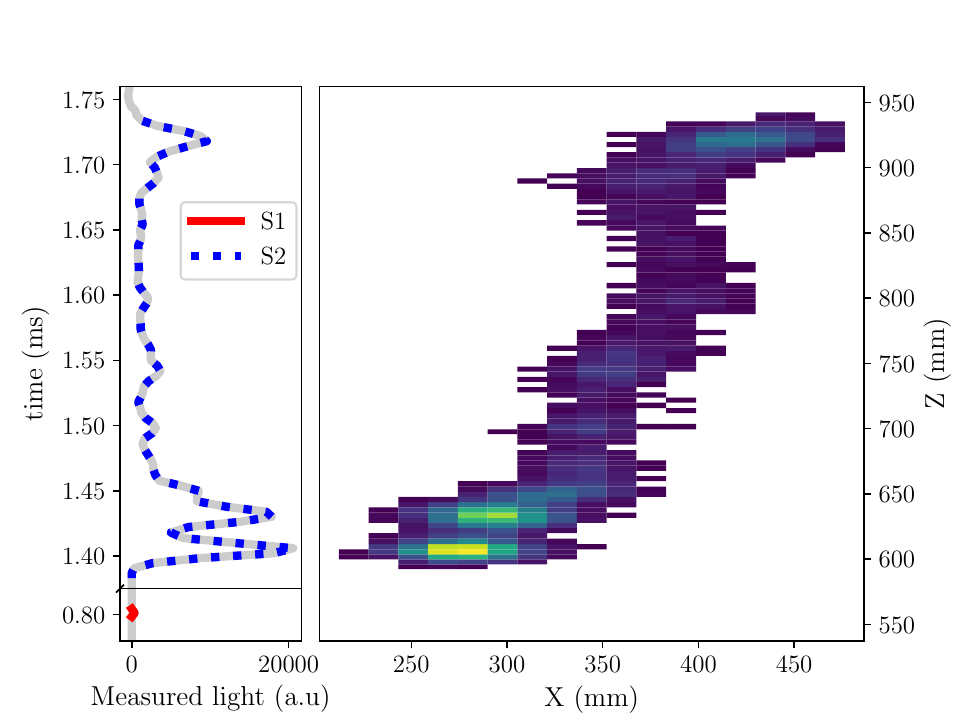}
\caption{Visualisation of a track collected in a LPR calibration run, from the initial raw waveforms from the PMTs (left), to the reconstructed energy deposits (hits) from the SiPMs collapsed into the X-Z plane (right). The prompt signal (S1) and the signal amplified through the electroluminescence region (S2) are highlighted. The hit energies are encoded in colour, with shades indicating higher energy.}\label{fig:S1S2}
\end{figure}

This production of photons so close to the tracking plane, combined with timestamping from the S1 and S2 signals, allows charged particle tracks to be reconstructed in 3D as `hits', individual energy deposits with \textit{X}, \textit{Y}, \textit{Z} positioning within the detector \cite{IC}. Here, \textit{X} and \textit{Y} are defined across the tracking plane and \textit{Z} along the length of the detector, as illustrated on the right of Figure \ref{fig:S1S2}.
The method for producing hits from the raw SiPM and PMT waveforms for each event relies on the successful identification of S1s and S2s. This is done using the summed PMT waveforms across the entire EP, enhancing the signal-to-noise ratio to allow for the successful identification of the small S1 signal, which provides the timestamp for its corresponding S2 signal, as seen in the left of Figure \ref{fig:S1S2}.

Once the S1 and S2 are identified, the drift time is defined as the time difference between them. Several analysis parameters are extracted for each S1 and S2 peak, including the width, height, and amount of light seen by both the SiPMs and PMTs, and are used to discard poorly reconstructed candidate events.

The S2 waveforms for each SiPM provide the $X$, $Y$ positioning for the resulting hits. 
The waveforms are integrated into \qty{4}{\micro\second} time-slices, each providing the amount of light detected per unit time, which is translated into a spatial coordinate along $Z$ using the known drift time.
The $X$ and $Y$ positions correspond to the location of each SiPM on the tracking plane, with all three coordinates defined in millimetres. 
This $X$, $Y$, $Z$ binning of the SiPM waveforms is the aforementioned hits, where the total energy of the event is reconstructed from the PMTs and distributed proportionally across each hit.
To mitigate coincident SiPM noise, a low-energy threshold is applied to each hit, and a clustering algorithm is used to reject nearest-neighbour groups with fewer than five hits \cite{N100_HE_eres}.


\section{Data selection \& topological reconstruction methods}\label{sec:data_and_topo}
\subsection{Data runs}\label{sec:ROI}
The data runs used in this analysis are dedicated high-energy calibration runs, in which an external source of $^{228}$Th is inserted into port 1A shown in Figure \ref{fig:detector}. The operating conditions across these runs can be found in Table \ref{tab:oper_params}. 
The decay chain of $^{228}$Th is dominated by \qty{2615}{keV} gamma rays from the decay of daughter $^{208}$Tl \cite{tl208_decay}. These gamma rays are energetic enough to induce pair production, which is used as the basis of this analysis due to the similarity of the $e^+e^-$ pairs to $0\nu\beta\beta$ events in both energy ($\mathcal{O}$(MeV)) and topology (two electron-like tracks, shared initial vertex). It should be noted that these events are limited in their comparison to genuine $0 \nu \beta \beta$ decay events due to their differing production kinematics and lower energy (\qty{1.6}{\mega\electronvolt} compared to \qty{2.5}{\mega\electronvolt}), the latter of which can have a significant impact on track topology.

\begin{table}[h]
\centering
\begin{tabular}{|l|l|}
\hline
\multicolumn{1}{|l|}{\textbf{Operational parameters}} & \textbf{Value}      \\ \hline
Pressure (bar)                               & 3.93       \\ \hline
Cathode voltage (kV)                          & 23         \\ \hline
Gate voltage (kV)                             & 8.8        \\ \hline
Drift field (V/cm)                           & 120        \\ \hline
Reduced EL Field (kV/cm/bar)             & 2.3        \\ \hline
\end{tabular}
\caption{Operating parameters during the runs of interest in this analysis.}
\label{tab:oper_params}
\end{table}

\subsection{Topological methodology}\label{sec:topoprocess}

\begin{figure}[ht]
    \hspace*{-2.7cm}\vspace{-1.5cm}\includegraphics[width=1.35\textwidth]{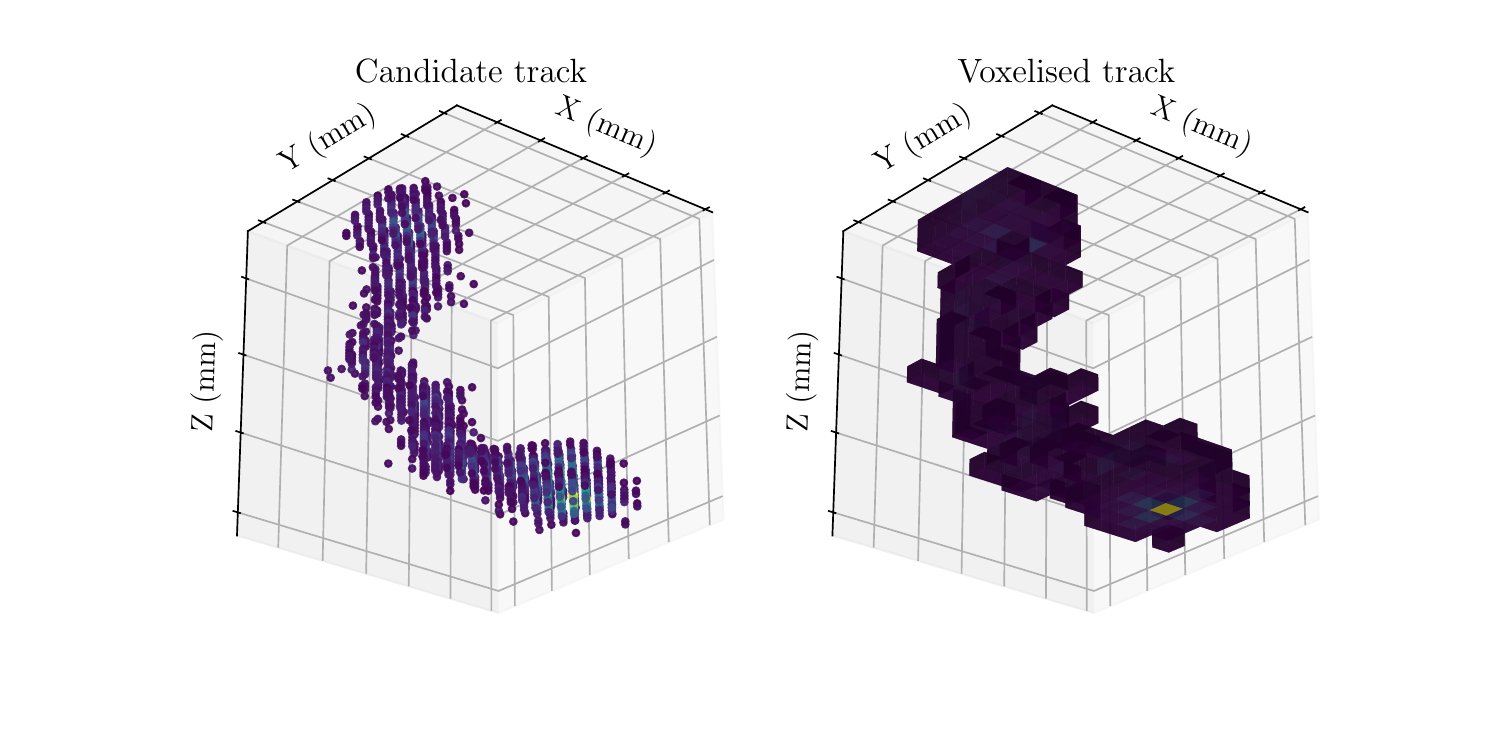}
    \caption{3D visualisation of the track shown in Figure \ref{fig:S1S2}, displaying the collection of independent hits (left) and the resulting voxelised track (right). The hit and voxel energies are encoded in the colour, with lighter shades indicating higher energy.}
    \label{fig:voxelisation}
\end{figure}

Hits are grouped into \textit{voxels}: grid-based 3D pixels that encapsulate multiple hits, as depicted in Figure \ref{fig:voxelisation}. A Breadth First Search algorithm \cite{bfs_skiena} is then applied across the voxels, based on their shared sides and vertices, to separate the voxels into connected groups, each of which forms a track.
From these tracks, topological information is extracted such as the extrema and length of each track, determined using Dijkstra's algorithm \cite{djikstra_algo}. Other parameters extracted explicitly from the tracks include the number of contained hits in each track, average track position, and track energy.

\begin{figure}[ht]
    \includegraphics[width=\textwidth]{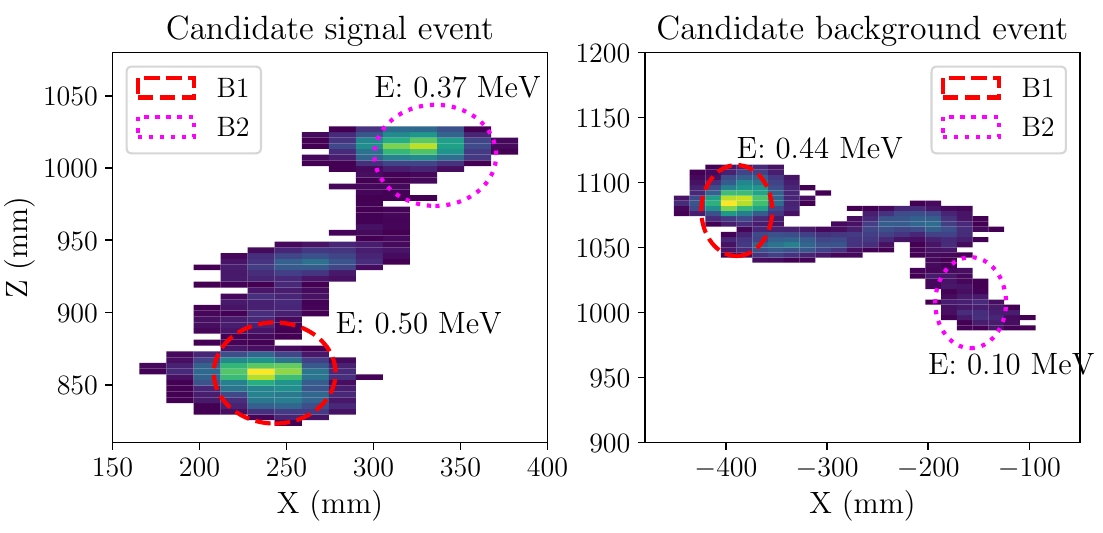}
    \caption{Demonstration of blob energy reconstruction for a signal (left) and background (right) event overlaid onto hits using \texttt{blob\_radius} of \qty{45}{mm}. The discrepancy in blob 2 energies is used to separate signal and background.}
    \label{fig:blob_energies}
\end{figure}

As charged particles pass through the detector and stop, they deposit most of their energy near the end of the track (at the Bragg peak) \cite{bragg_peak}. 
This allows for powerful discrimination between tracks with two Bragg peaks ($0\nu \beta \beta$ events) and those with only one (Compton-scattered electrons, X-ray emissions). To exploit this, all track extrema are identified as \textit{blobs}, and the energy encapsulated within a certain radius of each is defined as the \textit{blob energy}. The higher-energy blob is always defined as \textit{blob 1}, and the lower-energy blob as \textit{blob 2}, with the resulting discrepancy in blob 2 energies between one and two Bragg peak tracks providing strong discrimination power, as seen in Figure \ref{fig:blob_energies}.

\begin{figure}[H]
    \centering
    \hfill
    \begin{subfigure}{0.90\textwidth}
        \includegraphics[width=\textwidth]{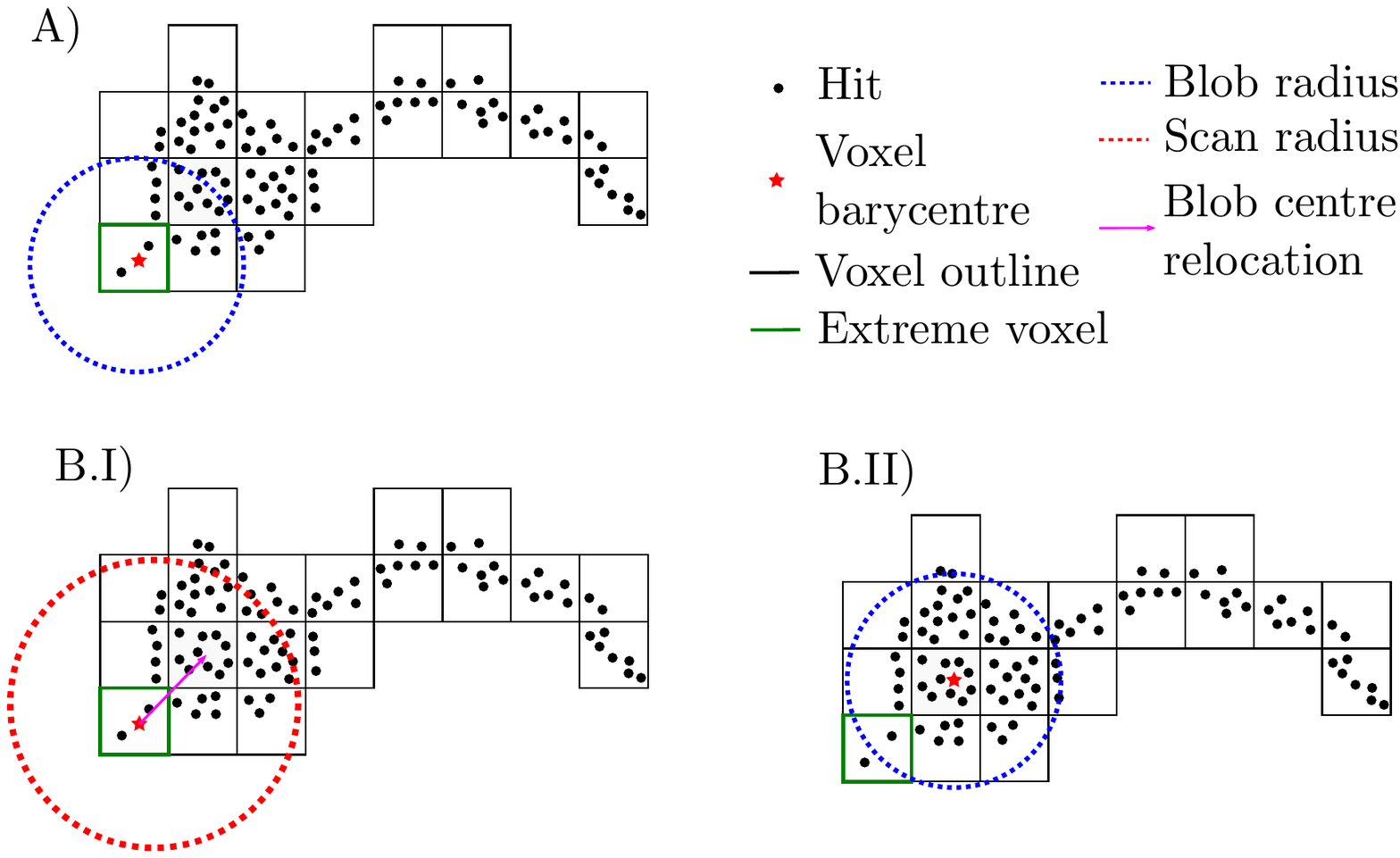}
    \end{subfigure}
    \caption{Visualisation of the blob reconstruction algorithms. A) shows the method used in prior analyses, in which the blob energy is calculated from hits within the blob radius (blue dashed line) centred on the most extreme voxel's barycentre (red star). B.I) and B.II) illustrate the new method. B.I) shows the first step, in which a scan radius (red dashed line) identifies the voxel that encapsulates the most energy within its own blob radius. The blob centre is then moved to that voxel (pink arrow). B.II) shows the blob energy being calculated from the updated voxel using the same method as A).}
    \label{fig:blob_reco}
\end{figure}

To accurately encapsulate the blob energy, the blob centre must be determined precisely. The method for determining the blob centre in prior analyses involved integrating around the track's extrema with a fixed radius. 
Reconstruction optimisation revealed that the aforementioned method for determining blob energies proved insufficient at capturing the blob energies accurately in a LPR context. This was understood to be due to an increase in blob size when compared to prior HPR analyses \cite{topology_NEW}, likely driven by the decreased pressure and reduced drift field resulting in increased diffusion and larger Bragg peaks as is discussed further in Section~\ref{ssec:pres_top}.
To resolve this issue, an improved algorithm is applied, which uses a larger \textit{scan radius} to collect all voxels within a given radius from the extrema. 
From these selected voxels, the one that encapsulates the most energy within a fixed radius around it (the \textit{blob radius}) is selected as the blob centre. A visualisation of this is seen in Figure \ref{fig:blob_reco} and the resulting improvement in topological performance in Figure~\ref{fig:FOM_visualiser}.

\section{Topological analysis}\label{sec:DEP_ana}
\subsection{Effects of pressure on topology}\label{ssec:pres_top}

The reduced operating pressure of \qty{4}{bar}, both relative to prior detectors in the NEXT programme and to NEXT-100's upcoming high-pressure run at $\sim$\qty{10}{bar}, raises new challenges and possible avenues of investigation.

The expectation is that decreased pressure allows longer tracks, which are easier to classify and study topologically. 
On the other hand, longer tracks result in reduced reconstruction efficiencies due to lower specific energy loss of the charged particle track, and an increased tendency to extend out of the fiducial volume. 
To better understand these effects at both 4 and \qty{10}{\bar}, an MC truth (direct GEANT4 energy depositions in the gas volume \cite{GEANT4}) study was performed to provide a general understanding of the effects of pressure on topology. 
The study simulates $0\nu \beta \beta$ events in the centre of the NEXT-100 chamber at both pressures, extracting the following characteristics of the resulting true electron tracks: 
\begin{itemize}
    \item Maximum extent (mm): the maximum distance between any two parts of the track,
    \item True length (mm): the total length along the track,
    \item Blob size (mm): the radius required to encapsulate \qty{400}{keV} of energy from the track extrema,
    \item Specific energy loss (keV/mm): the energy loss per unit of distance along the track.
\end{itemize}

The median values for these topological parameters extracted across the MC sample can be seen in Table \ref{tab:pressure}. 
\begin{table}[ht]
\centering
\begin{tabular}{|l|l|l|}
\hline
\textbf{Pressure}             & \textbf{4 bar}       & \textbf{10 bar}      \\ \hline
Maximum extent (mm)           & 280                  & 108                  \\ \hline
True length (mm)              & 877                  & 337                  \\ \hline
Blob size (mm)                & 28                   & 12                   \\ \hline
Specific energy loss (keV/mm) & 2.9                  & 9.3                  \\ \hline
\end{tabular}
\caption{Comparison of topological characteristics for $0 \nu \beta \beta$ tracks at differing pressures from the Monte Carlo truth study.}
\label{tab:pressure}
\end{table}

It is observed that the true tracks travel $\sim$2.6 times the distance at \qty{4}{bar} compared to \qty{10}{bar}, with an increased blob size (by a factor of $\sim$2.3) and a decreased specific energy loss (by a factor $\sim$3). 
As these are true electron tracks prior to detector simulation, diffusion effects are not included and as such the blob sizes found here will be smaller than those found in data. Compared to NEXT-White at \qty{10}{bar}, the longitudinal and transverse diffusion are expected to increase from $\sim$\qty{0.3}{} and $\sim$\qty{1.1}{mm\,/\sqrt{\text{cm}}} to $\sim$\qty{0.55}{} and $\sim$\qty{1.88}{mm\,/\sqrt{\text{cm}}} respectively for NEXT-100 in LPR conditions \cite{NEW_diffusion}. 
In line with these findings, tracks in the NEXT-100 LPR run were observed to be longer, with less deposited energy per unit distance and larger blobs when compared to NEXT-White at \qty{10}{bar}, with results for this comparison found in Appendix \ref{app:blob_radii}.

\subsection{Double escape peak analysis}\label{ssec:dep_ana_proper}
As mentioned in Section \ref{sec:ROI}, this analysis relies on $e^+e^-$ pairs produced via pair production from \qty{2615}{keV} gamma rays, focusing specifically on events where the resulting positron annihilates, releasing two \qty{511}{keV} photons that escape the detector without interacting. This produces a distinct peak in the energy spectrum at $\sim$\qty{1593}{keV}, known as the \textit{double escape peak} (DEP).
    
Double escape peak events act as an excellent benchmark for background rejection, as their tracks contain two high-energy blobs while background (single electron) tracks contain only one high-energy blob, mimicking a real search for $0 \nu \beta \beta$ events.
To ensure the sample studied consists of well-reconstructed events relevant to this analysis, a set of data quality selections is applied, in the following order: 
\begin{itemize}

    \item to isolate events around the DEP, a region of interest cut is applied to the track energies, selecting events within a small window around the peak (1.4-\qty{1.8}{MeV}),

    \item hits reconstructed near the edge of the detector suffer from reduced reconstruction quality. To resolve this, a fiducial cut is applied, requiring $\qty{20}{mm} < Z < \qty{1170}{mm}$ and radius $R < \qty{450}{mm}$, discarding any tracks with hits outside this region, 

    \item secondary emissions from DEP events (via Bremsstrahlung and Compton scattering) can produce additional reconstructed tracks, drawing energy away from the main track and distorting its blob energies; therefore a one-track cut is applied, removing any events with more than one reconstructed track, 

    \item overlapping blobs introduce difficulties in determining the energy attributable to each blob, so a blob-overlap cut is applied, discarding any tracks whose unique blobs physically overlap.

\end{itemize}



\begin{figure}[ht]
    \centering
    \hspace{-1.3em}
    \makebox[\textwidth][c]{%
        \begin{subfigure}{0.55\textwidth}
            \includegraphics[width=\textwidth]{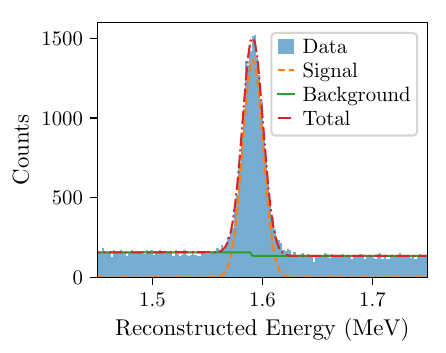}
        \end{subfigure}%
        \begin{subfigure}{0.445\textwidth}
            \includegraphics[trim={1.45cm 0 0 0}, clip, width=\textwidth]{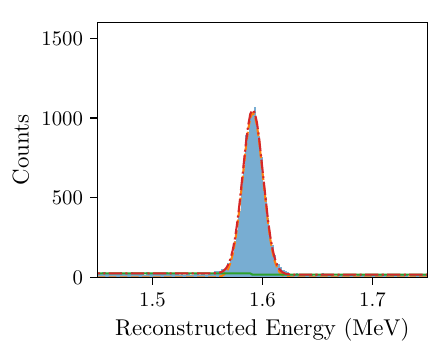}
        \end{subfigure}%
    }
    \caption{Double escape peak energy spectra before blob-2 cuts (left) and at the optimal blob-2 cut (right), with the exponential and step function (background) fit in solid green, the Gaussian (signal) fit in dashed orange, and the combined fit with red dash-dot hatching.}
    \label{fig:FOM_plots}
\end{figure}

These cuts and the resulting efficiencies are provided in Appendix \ref{app:cut_flow}.
To determine the efficiency of blob discrimination for background removal, an unbinned maximum likelihood fit is applied to the reconstructed energy spectrum around the DEP. The background is modelled as the sum of an exponential, describing the Compton background, and a step function centred on the peak, accounting for DEP events with partial energy loss. The DEP signal is modelled using a Gaussian, with both fits and their combination shown in Figure \ref{fig:FOM_plots}. The number of signal and background events are determined over the signal region using the fit within a $\pm3\sigma$ window around the Gaussian mean. 
Cuts are then applied iteratively on blob 2 energy, scanning from 0 to \qty{0.6}{MeV} in steps of \qty{0.0167}{MeV}, with the fit reapplied after each cut to recalculate the number of signal and background events. These estimates are then used to calculate a \textit{Figure of Merit} (FOM) as follows:

\begin{equation}
    FOM = \frac{e^i}{\sqrt{b^i}},
\end{equation}
where,
\begin{equation}
    e^i = \frac{N_s^i}{N_s^0},\
    b^i = \frac{N_b^i}{N_b^0}, 
\end{equation}
with $N_{s,(b)}^i$ being the number of signal (background) events after the \textit{i}'th cut, and $N_{s,(b)}^0$ being before any cuts are applied.


\begin{figure}[ht]
    \hspace{-1.5em}
    \includegraphics[width=1.05\textwidth]{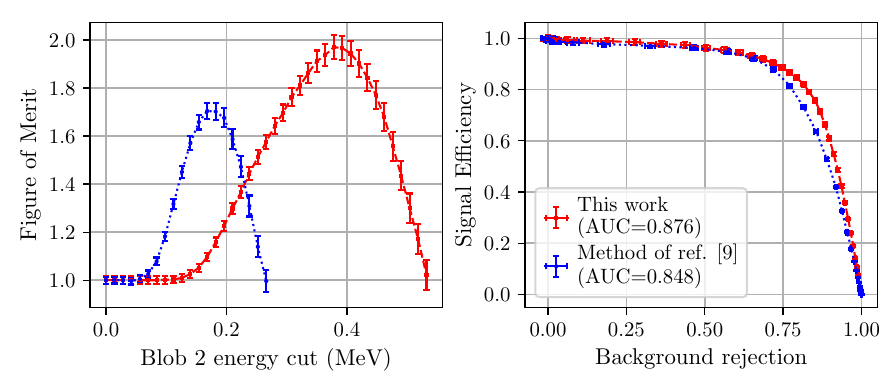}
    \caption{The figure of merit (left) and signal efficiency against background rejection (right) for NEXT-100 utilising the blob reconstruction algorithm discussed in this work (in red) and the prior algorithm from \cite{topology_NEW} (in blue). The NEXT-100 results shown here are implemented with optimised parameters: \texttt{blob\_radius}: \qty{45}{mm}, \texttt{scan\_radius}: \qty{60}{mm}, \texttt{voxel\_size}: \qty{15}{mm}, with the prior algorithm excluding the \texttt{scan\_radius} parameter.}
    \label{fig:FOM_visualiser}
\end{figure}

The reconstruction and topological parameters are then optimised using the FOM and selection efficiencies as metrics, with a parameter scan applied across three key parameters relevant to the improved blob centre algorithm described in Section \ref{sec:topoprocess}:

\begin{itemize}
    \item \texttt{blob\_radius}; the radius within which the blob's energy is encapsulated,
    \item \texttt{scan\_radius}; the radius from the track's endpoint defining which voxels to consider as the possible centre,
    \item \texttt{voxel\_size}; the size of the square voxels. 
\end{itemize}
The optimal parameters were determined to be: \texttt{blob\_radius} \qty{45}{mm}, \texttt{scan\_radius} \qty{60}{mm}, and a \texttt{voxel\_size} of \qty{15}{mm} (the corresponding parameter scans can be seen in Appendix \ref{app:pscans}). 
These parameters provide a maximal FOM of $1.97\pm 0.05\,\text{(stat.)}\,^{+0.09}_{-0.10}\,\text{(syst.)}$ at a blob cut of 0.38 MeV, with a signal efficiency of $75.6\pm 1.9\,\text{(stat.)}\,^{+3.4}_{-4.1}\,\text{(syst.)}\,\text{\%}$ and background acceptance of $14.7\pm 0.4\,\text{(stat.)}\,^{+0.7}_{-0.8}\,\text{(syst.)}\,\text{\%}$ as seen in Figure \ref{fig:FOM_visualiser}. 
The statistical errors are dominated by the fit uncertainties on the number of signal and background events. Consequently, neglecting the correlation between the samples before and after the blob 2 energy cut yields a conservative estimate for the statistical error. 
The systematics are estimated by rerunning the FOM analysis with reconstruction parameters varied one at a time around the nominal (optimised) value, keeping all other parameters fixed. 
The tested range for each parameter was chosen to span values considered plausible alternatives to the optimised nominal, informed through simple parameter scans such as those shown in Appendix \ref{app:pscans}.
The maximal deviation from nominal across the tested values is used as the uncertainty for that parameter, with asymmetric deviations retained where upward and downward deviations differed appreciably.
The individual contributions are then combined in quadrature, including an extrapolation of systematics for geometrical corrections from prior works \cite{Kr_NEW}.
A comparison to previous analyses for the NEXT-White detector at HPR can be seen in Table \ref{tab:FOMs}. 

\begin{table}[ht]
\centering
\begin{tabular}{l|c|c|c|}
\cline{2-4}& \textbf{FOM} & \textbf{\begin{tabular}[c]{@{}c@{}}Signal efficiency \\ (\%)\end{tabular}} & \textbf{\begin{tabular}[c]{@{}c@{}}Background acceptance \\ (\%)\end{tabular}} \\ \hline

\multicolumn{1}{|l|}{\begin{tabular}[c]{@{}l@{}}\ NEXT-100 LPR Classical \\ (Method of ref. \cite{topology_NEW})\end{tabular}} & 1.70$^{+0.08}_{-0.09}$         & 81.5$^{+3.9}_{-4.6}$                                                                       & 22.9$^{+1.1}_{-1.3}$                                                                           \\ \hline
\multicolumn{1}{|l|}{\begin{tabular}[c]{@{}l@{}}\ NEXT-100 LPR Classical \\ (This work)\end{tabular}} & 1.97$^{+0.10}_{-0.11}$         & 75.6$^{+3.9}_{-4.5}$                                                                       & 14.7$^{+0.8}_{-0.9}$                                                                           \\ \hline
\multicolumn{1}{|l|}{\begin{tabular}[c]{@{}l@{}}NEXT-White \\ Classical \cite{topology_NEW}\end{tabular}}      & 1.58$\pm$0.04         & 71.6$\pm$1.5                                                                      & 20.6$\pm$0.4                                                                           \\ \hline
    \multicolumn{1}{|l|}{\begin{tabular}[c]{@{}l@{}}NEXT-White \\ RL-Deconvolution \cite{deconv_NEW}\end{tabular}}      & 2.94$\pm$0.28        & 56.6$\pm$2.2                                                                      & 3.7$\pm$0.7                                                                            \\ \hline
    \multicolumn{1}{|l|}{\begin{tabular}[c]{@{}l@{}}NEXT-White \\ CNN \cite{CNN_NEW}\end{tabular}}            & 2.2          & 70                                                                         & 10                                                                             \\ \hline
\end{tabular}

\caption{Comparison of signal figure of merit, signal efficiency and background acceptance for this analysis with previous NEXT results \cite{topology_NEW, deconv_NEW, CNN_NEW}. The baseline cut-based method is described here as the `Classical' method in line with previous analyses, and the uncertainty for the NEXT-100 LPR runs is provided as the systematic and statistical uncertainties combined in quadrature.}
\label{tab:FOMs}
\end{table}

\section{Discussion}\label{sec:discussion}
The methodology used throughout this analysis demonstrates the strengths of topological discrimination in the context of gaseous detectors, and the robustness of this technology with respect to variable operating pressures.

The signal efficiency ($75.6\pm 1.9\,\text{(stat.)}\,^{+3.4}_{-4.1}\,\text{(syst.)}\,\text{\%}$) and background acceptance ($14.7\pm 0.4\,\text{(stat.)}\,^{+0.7}_{-0.8}\,\text{(syst.)}\,\text{\%}$) result in a figure of merit ($1.97\pm 0.05\,\text{(stat.)}\,^{+0.09}_{-0.10}\,\text{(syst.)}$) yielding an improvement over the cut-based analysis of NEXT-White at \qty{10}{bar} (Table \ref{tab:FOMs}) \cite{topology_NEW}.

Attributing this improvement between detector, algorithmic, and pressure effects is non-trivial. Some insight is gained by applying the previous fixed-centre blob method to the same LPR dataset, which yields a maximal figure of merit of $1.70 \pm 0.10\,\text{(stat.)}\,^{+0.08}_{-0.09}\,\text{(syst.)}$ (Figure~\ref{fig:FOM_visualiser}), a modest improvement over the cut-based NEXT-White result.
This suggests that the topological benefit of longer tracks at low pressure is partly offset by the accompanying diffusion and blob growth, with the full benefit recovered once the reconstruction is adapted to these conditions. Whether the improved blob-centring algorithm would yield a similar gain at \qty{10}{bar}, where blobs are smaller and denser, remains to be tested in the coming HPR runs.

A significant difference in the peak location (\qty{0.18}{MeV} to \qty{0.38}{MeV}) between the prior blob encapsulation method \cite{topology_NEW} and the method presented in this work can be seen in Figure \ref{fig:FOM_visualiser}. This is understood to result from the improved energy encapsulation of the newer method. Due to the blob centres being relocated away from the track extrema, blob energies are inherently larger in both signal and background events, shifting the optimal blob 2 cut to higher values. 

The MC truth study of Section~\ref{sec:DEP_ana} (Table~\ref{tab:pressure}) indicates that reducing the pressure from 10 to \qty{4}{bar} lengthens tracks by a factor of $\sim$2-3 while reducing specific energy loss by a factor of $\sim$3.
The track length increase is observed in data (Appendix~\ref{ssec:track_length}); at fixed track energy, this directly implies the corresponding reduction in specific energy loss. This reduction allows for increased track splitting and reduced single-track efficiencies (Section~\ref{sec:DEP_ana}). Longer tracks improve the separation between the track ends, aiding blob discrimination. However, the blob size also grows, by a factor of $\sim$2.3 in MC and nearly a factor of 2 in data (Appendix~\ref{ssec:blob_rad}), alongside increased diffusion \cite{NEW_diffusion}. Together these effects produce broader, less dense tracks that degrade conventional blob reconstruction, motivating the blob-centring method of Section~\ref{sec:topoprocess} and the re-optimised reconstruction parameters.

It should be noted that operational differences between the two detectors (NEXT-White and NEXT-100), such as the differing SiPM pitch, applied drift field, and source port location relative to the tracking plane, complicate a direct comparison. Further studies into the effect of pitch on topological reconstruction are ongoing.

Longer tracks are also more likely to extend beyond the fiducial volume of the detector, reducing the fiducial cut efficiency and thereby requiring more runtime to achieve the same number of events passing all selection cuts. This efficiency is expected to recover in higher-pressure runs.

The cut-based analysis presented here is considered the `baseline' method for NEXT topological analysis.
It was shown for NEXT-White that more sophisticated methods for reconstruction (such as Richardson-Lucy deconvolution \cite{deconv_NEW}) and classification (such as convolutional neural networks \cite{CNN_NEW}) increase the topological power over the baseline method as seen in Table \ref{tab:FOMs}. 
This is assumed to hold for NEXT-100, where these more powerful methods will be applied across the coming HPR runs.

This first demonstration of topological selection with the NEXT-100 detector, even at low pressure, highlights the detector's versatility and performance in the early stages of its scientific programme, providing confidence in its capabilities for future scientific runs.

The LPR runs offer a unique opportunity for further study into track information that is more difficult to resolve at higher pressure, such as the emission angle between the two outgoing electron-like particles. Improved understanding of how to extract this information at differing operating pressures could enable greater sensitivity to the underlying decay mechanism of $0 \nu \beta \beta$ \cite{angle_by_mass}, motivating continued interest in lower-pressure TPCs \cite{atmo_tpc}.

The performance achieved by the NEXT-100 detector at LPR is comparable to that obtained at HPR in NEXT-White in all key areas (topological discrimination, stability throughout calibration \cite{N100_kr}, and energy resolution \cite{N100_HE_eres}). This suggests that future experiments could begin operation with xenon quantities significantly below their designed capacity and progressively increase both the xenon mass and operating pressure as additional enriched xenon becomes available.
Since the procurement of large quantities of enriched xenon can take several years, such a staged deployment strategy could significantly accelerate the physics programme. This flexibility is a unique advantage of gaseous TPCs and strengthens their case as a competitive technology for next-generation tonne-scale neutrinoless double beta decay searches.

\section{Conclusion}\label{sec:conclusion}

In summary, this work has demonstrated the conventional cut-based topological capabilities of NEXT-100, establishing the baseline for the detector's background rejection at LPR with a figure of merit of $1.97\pm 0.05\,\text{(stat.)}\,^{+0.09}_{-0.10}\,\text{(syst.)}$, corresponding to a signal efficiency and background acceptance of $75.6\pm 1.9\,\text{(stat.)}\,^{+3.4}_{-4.1}\,\text{(syst.)}\,\text{\%}$ and $14.7\pm 0.4\,\text{(stat.)}\,^{+0.7}_{-0.8}\,\text{(syst.)}\,\text{\%}$ respectively around the $^{208}$Tl double escape peak. 
The results and methodology shown here will act as a foundation for further topological investigations in the NEXT programme throughout the coming HPR run, applying methods previously demonstrated in NEXT-White, such as Richardson-Lucy deconvolution and machine-learning processes, to improve upon the cut-based method presented in this work.

\acknowledgments
The NEXT Collaboration acknowledges support from the following agencies and institutions: the European Research Council (ERC) under Grant Agreements No. 951281-BOLD and 101039048-GanESS; the European Union’s Framework Programme for Research and Innovation Horizon 2020 (2014–2020) under Grant Agreement No. 860881-HIDDeN; the MCIN/AEI of Spain and ERDF A way of making Europe under grants PID2021-125475NB and RTI2018-095979, and the Severo Ochoa and María de Maeztu Program grants CEX2023-001292-S, CEX2023-001318-M and CEX2018-000867-S; the Generalitat Valenciana of Spain under grants PROMETEO/2021/087, ASFAE/2022/028, ASFAE/2022/029, CISEJI/2023/27 and CIDEXG/2023/16; the Department of Education of the Basque Government of Spain under the predoctoral training program of non-doctoral research personnel; the Spanish la Caixa Foundation (ID 100010434) under fellowship code LCF/BQ/PI22/11910019; the Portuguese FCT under project UID/FIS/04559/2020 to fund the activities of LIBPhys-UC; the Israel Science Foundation (ISF) under grant 1223/21; the Pazy Foundation (Israel) under grants 310/22, 315/19 and 465; the US Department of Energy under contracts number DE-AC02-06CH11357 (Argonne National Laboratory), DE-AC02-07CH11359 (Fermi National Accelerator Laboratory), DE-FG02-13ER42020 (Texas A\&M), DE-SC0019054 (Texas Arlington) and DE-SC0019223 (Texas Arlington); the US National Science Foundation under award number NSF CHE 2004111; the Robert A Welch Foundation under award number Y-2031-20200401. Finally, we are grateful to the Laboratorio Subterráneo de Canfranc for hosting and supporting the NEXT experiment.

\appendix
\section{Figure of Merit optimisation}\label{app:pscans}

Two separate parameter scans were completed to determine optimal parameters for topological reconstruction. The first scan independently varied \texttt{blob\_radius} and \texttt{scan\_radius} with a fixed \texttt{voxel\_size}, while the second treated both radii as a linked pair alongside \texttt{voxel\_size}. These scans resulted in a choice of parameters that balances maximising the figure of merit while ensuring the selection cuts described in Section \ref{sec:DEP_ana} retained acceptable efficiencies.
Both scans were applied on subsets of the high-energy calibration runs to reduce runtime; consequently, the metrics do not align exactly with the final result presented in this analysis. 
\newpage
\subsection{Scan 1}

\begin{figure}[ht]
    \centering
    \begin{subfigure}{0.49\textwidth}
        \includegraphics[width=\textwidth]{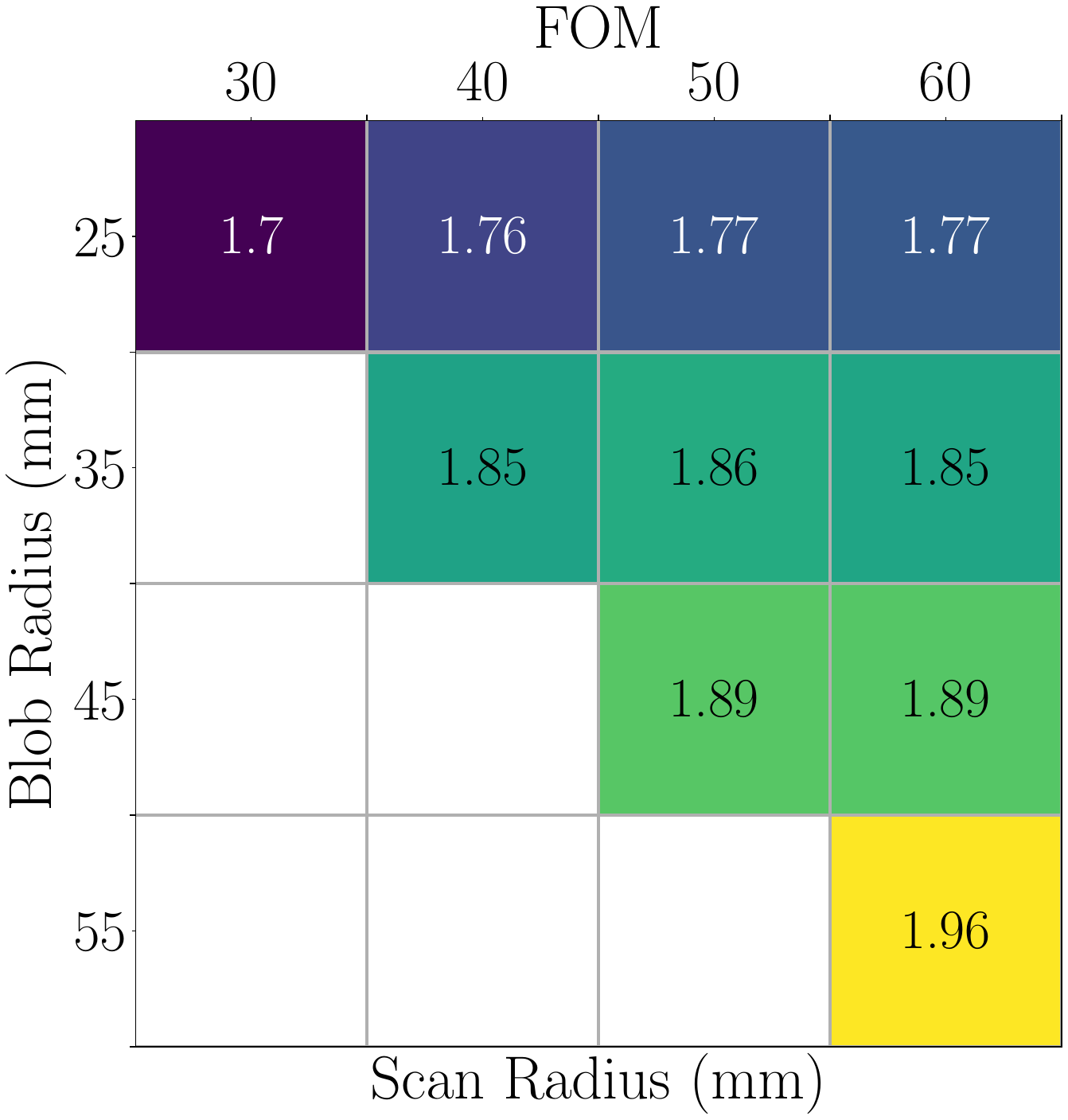}
    \end{subfigure}
    \hfill
    \begin{subfigure}{0.49\textwidth}
        \includegraphics[width=\textwidth]{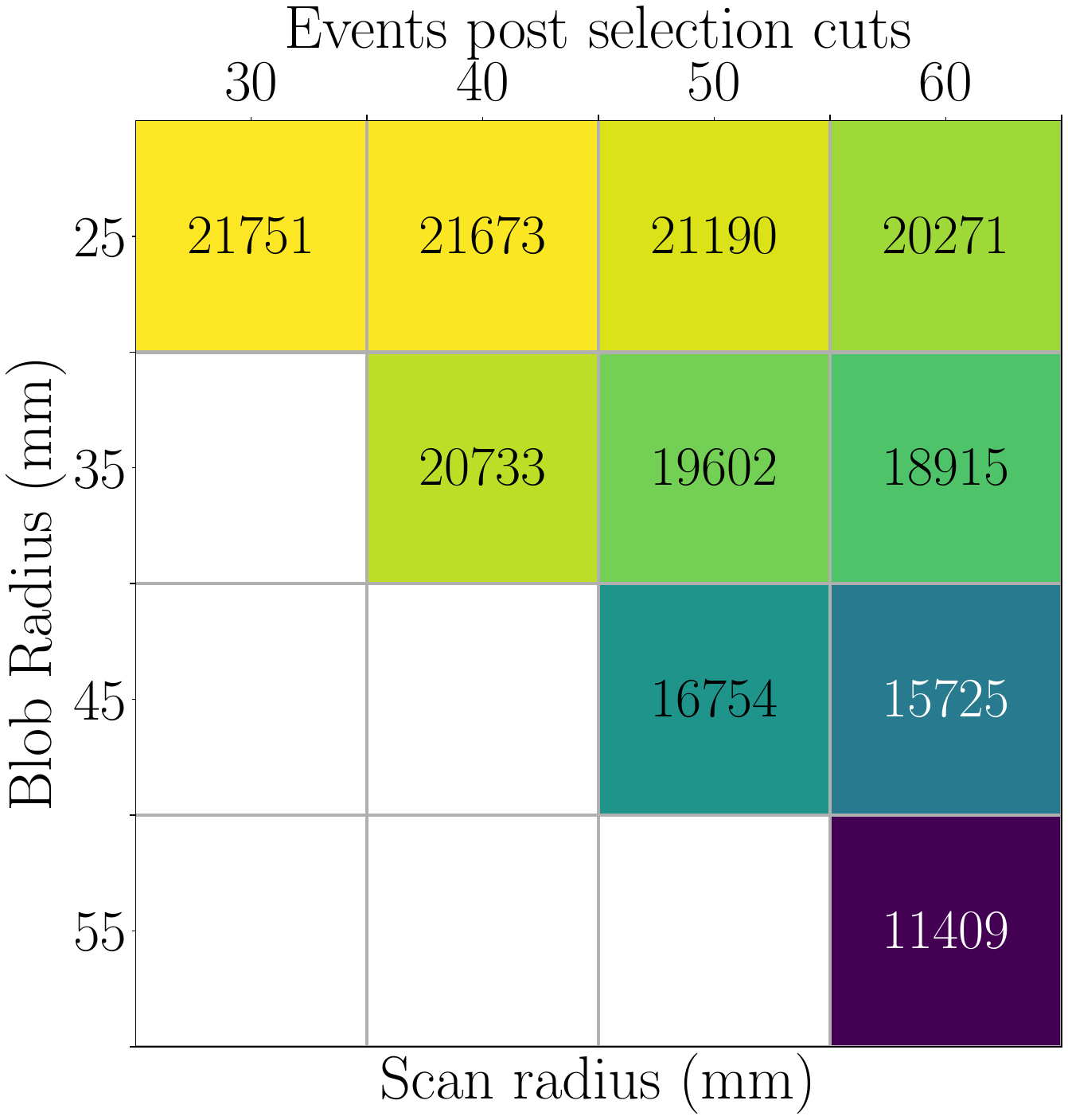}
    \end{subfigure}
    \begin{subfigure}{0.49\textwidth}
        \includegraphics[width=\textwidth]{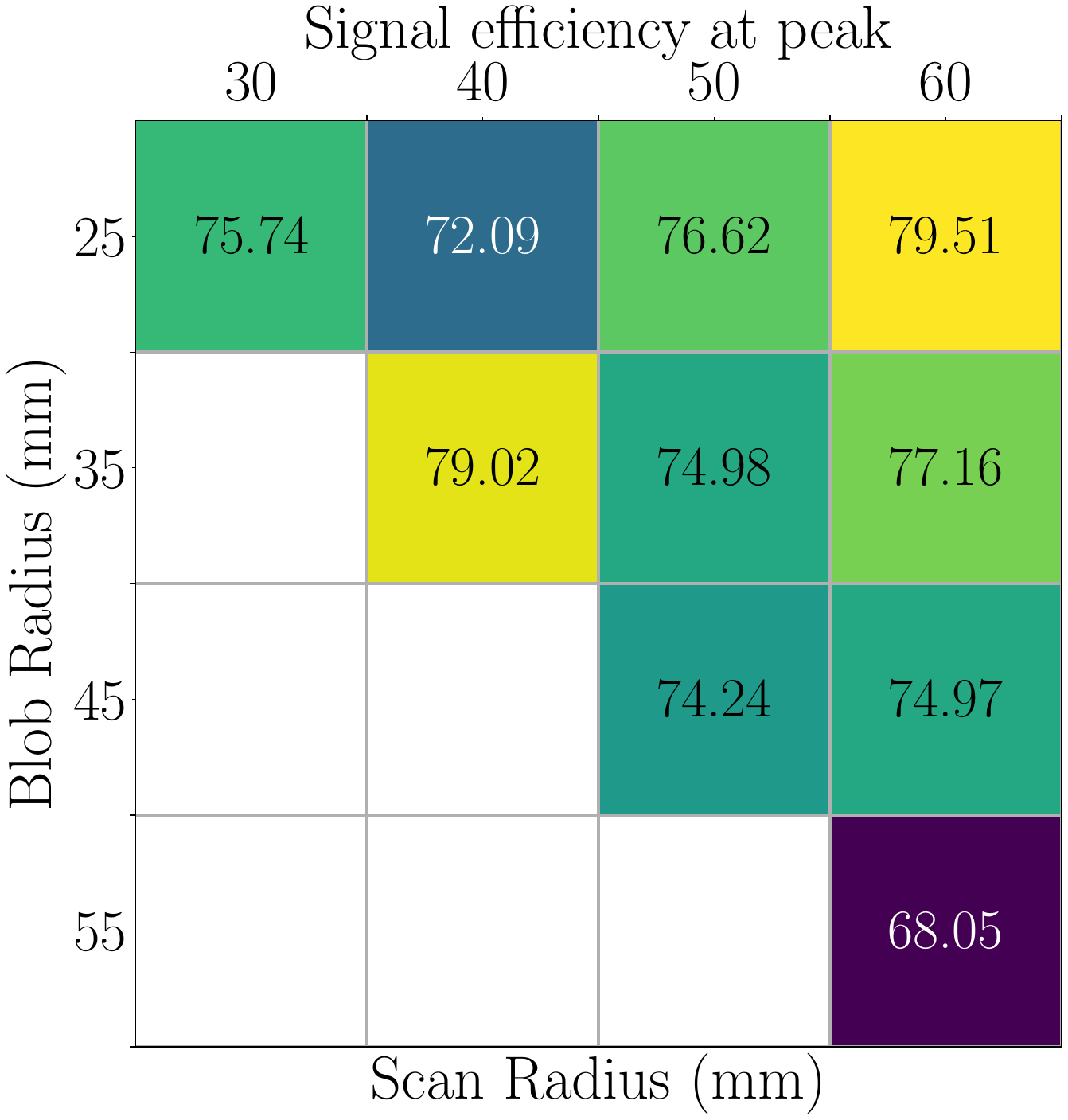}
    \end{subfigure}
    \begin{subfigure}{0.49\textwidth}
        \includegraphics[width=\textwidth]{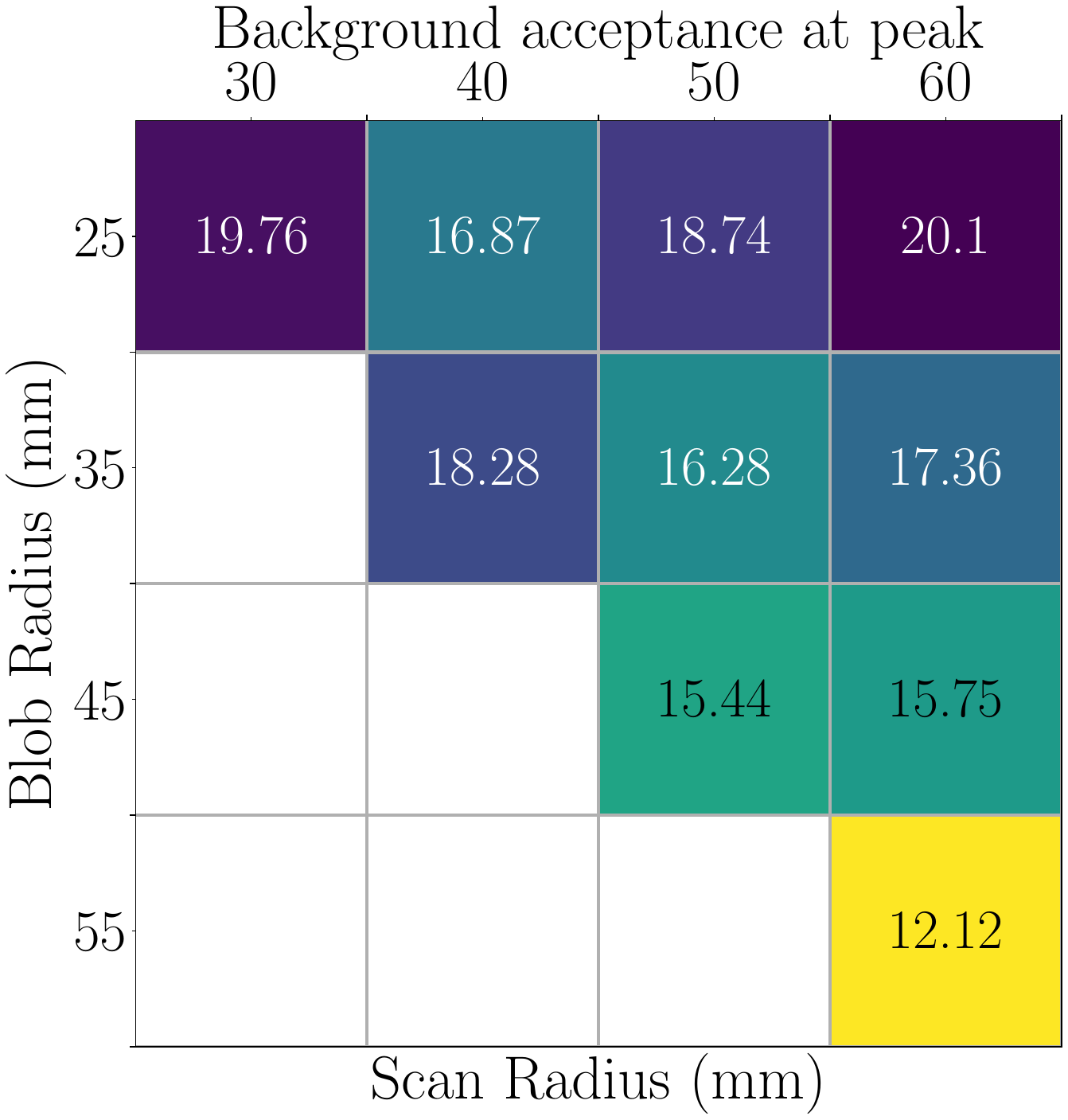}
    \end{subfigure}
    \caption{Parameter scan across blob and scan radius, with each quadrant illustrating differing performance metrics. Top left: maximal figure of merit (FOM). Top right: total number of events after selection cuts are applied as described in Section \ref{sec:DEP_ana}. Bottom left: signal efficiency at maximal FOM. Bottom right: background acceptance at maximal FOM.}
\end{figure}
\newpage
\subsection{Scan 2}

\begin{figure}[h]
    \centering
    \begin{subfigure}{0.49\textwidth}
        \includegraphics[width=\textwidth]{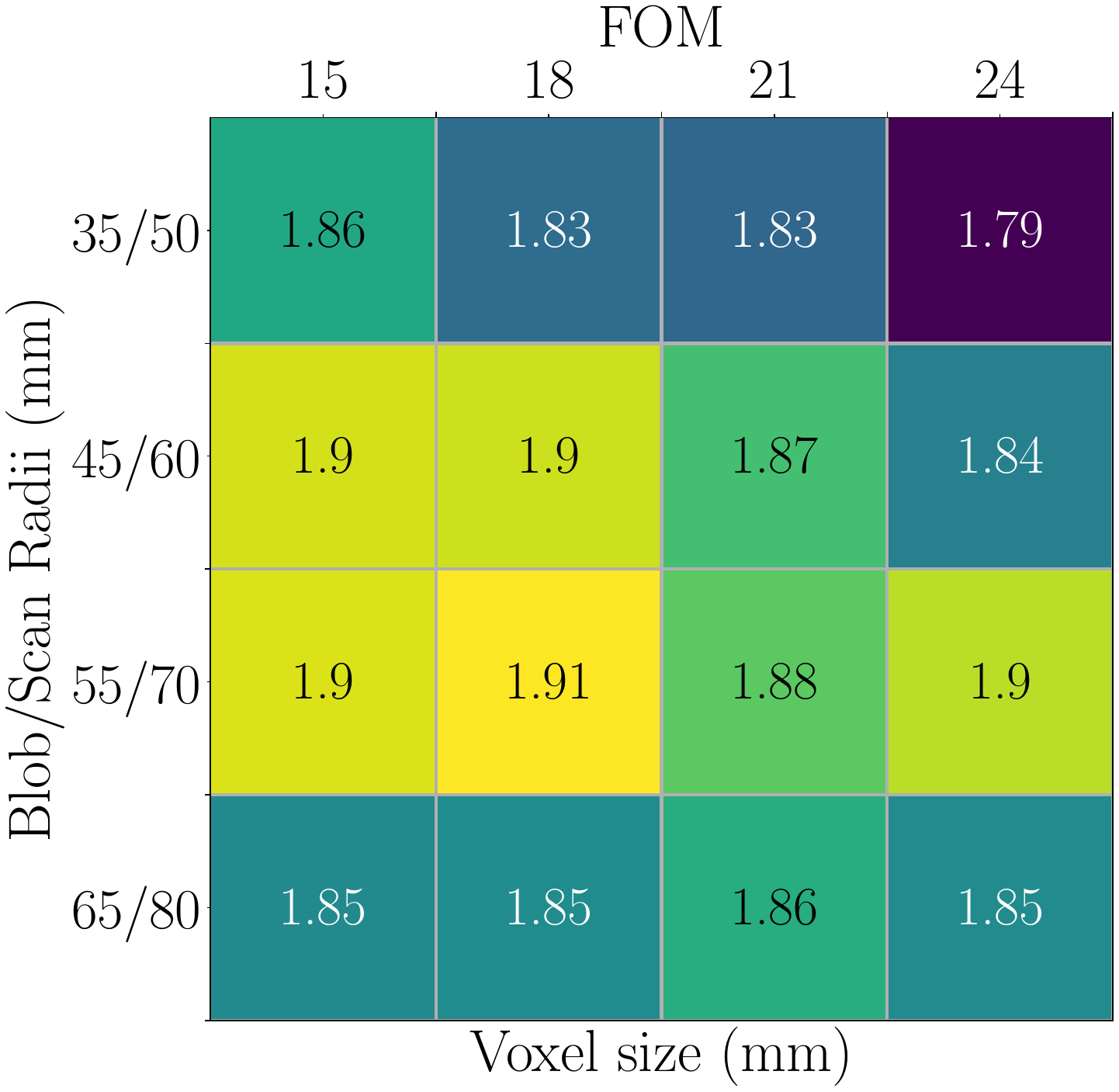}
    \end{subfigure}
    \hfill
    \begin{subfigure}{0.49\textwidth}
        \includegraphics[width=\textwidth]{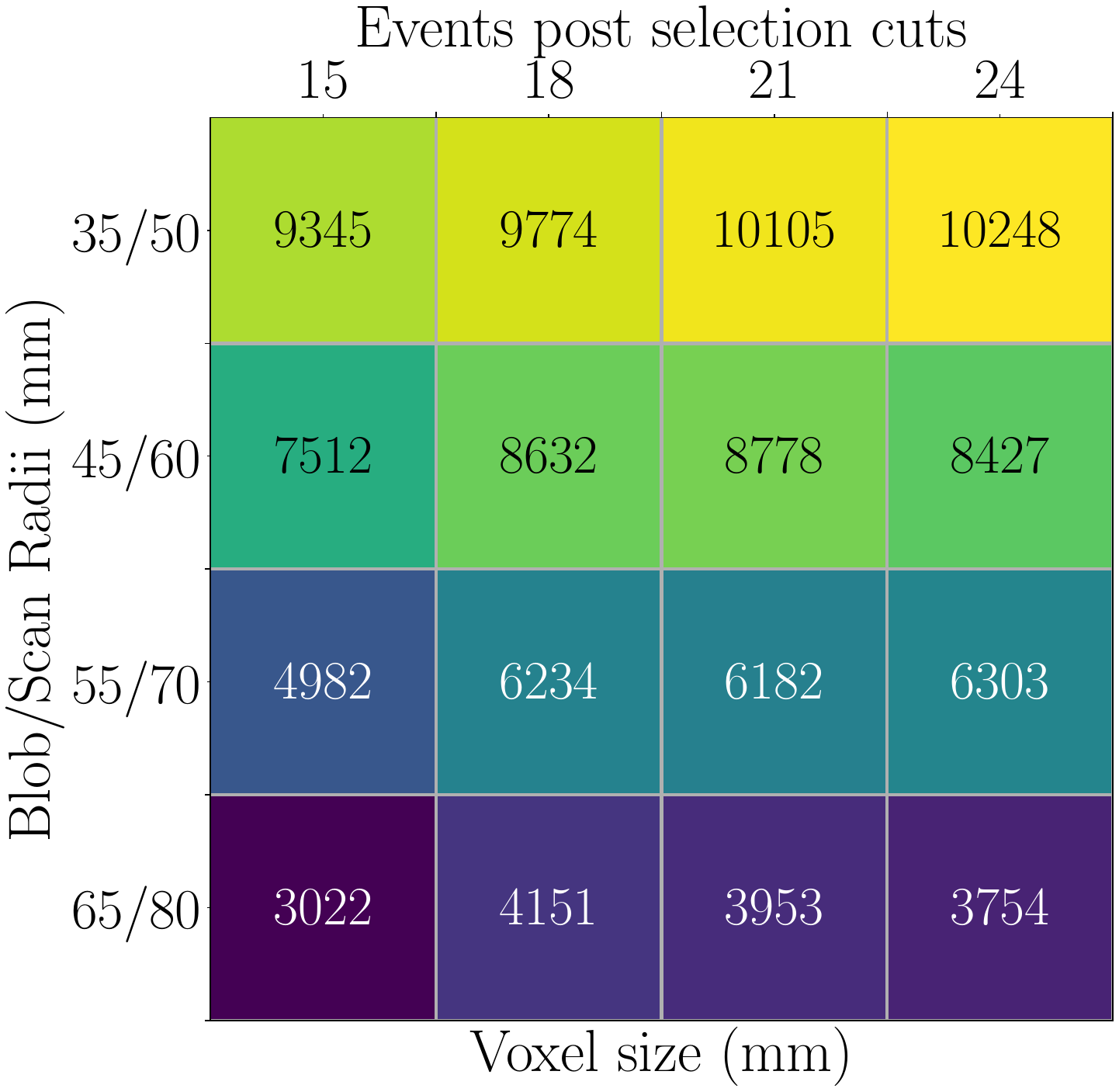}
    \end{subfigure}
    \begin{subfigure}{0.49\textwidth}
        \includegraphics[width=\textwidth]{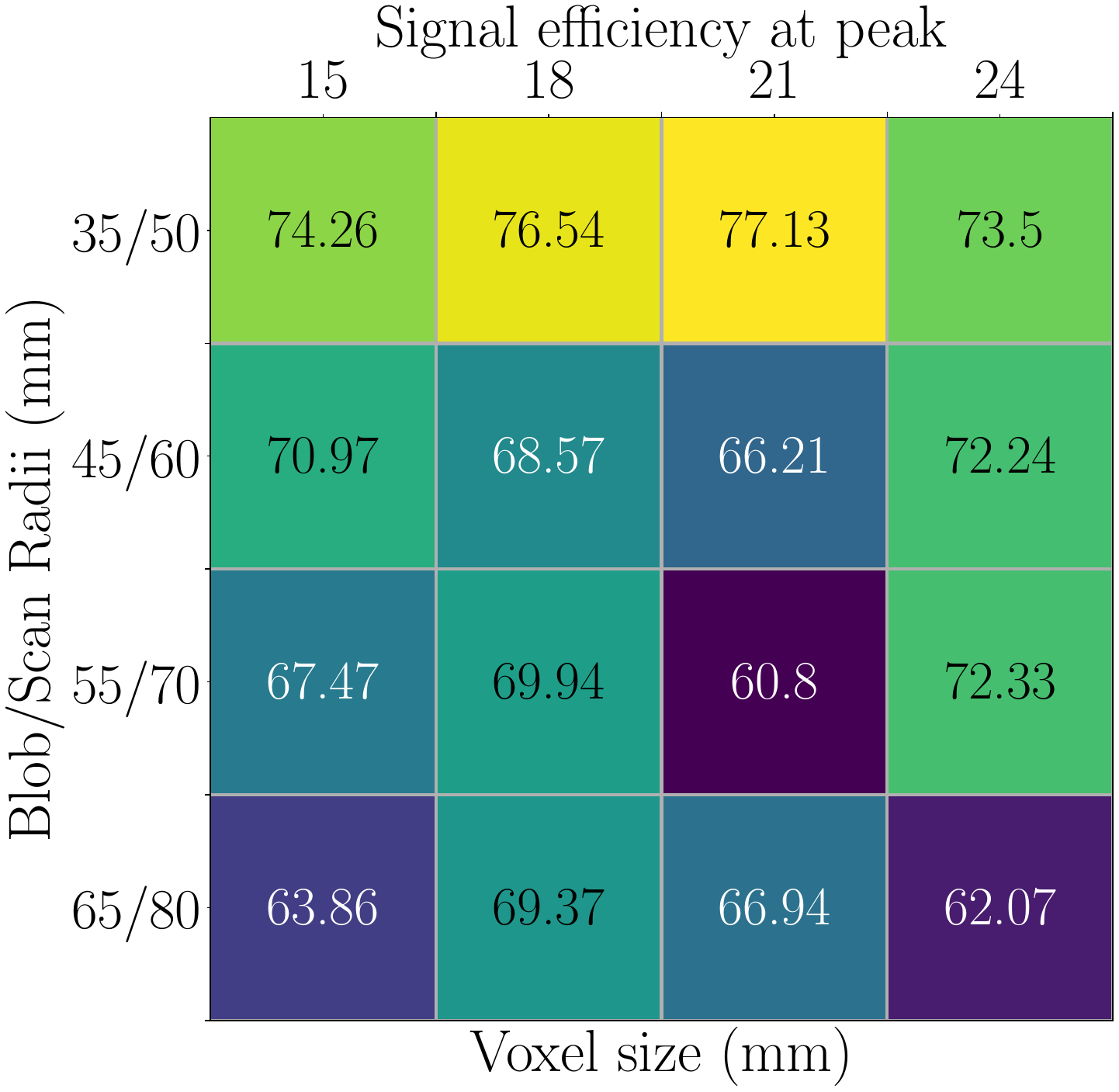}
    \end{subfigure}
    \begin{subfigure}{0.49\textwidth}
        \includegraphics[width=\textwidth]{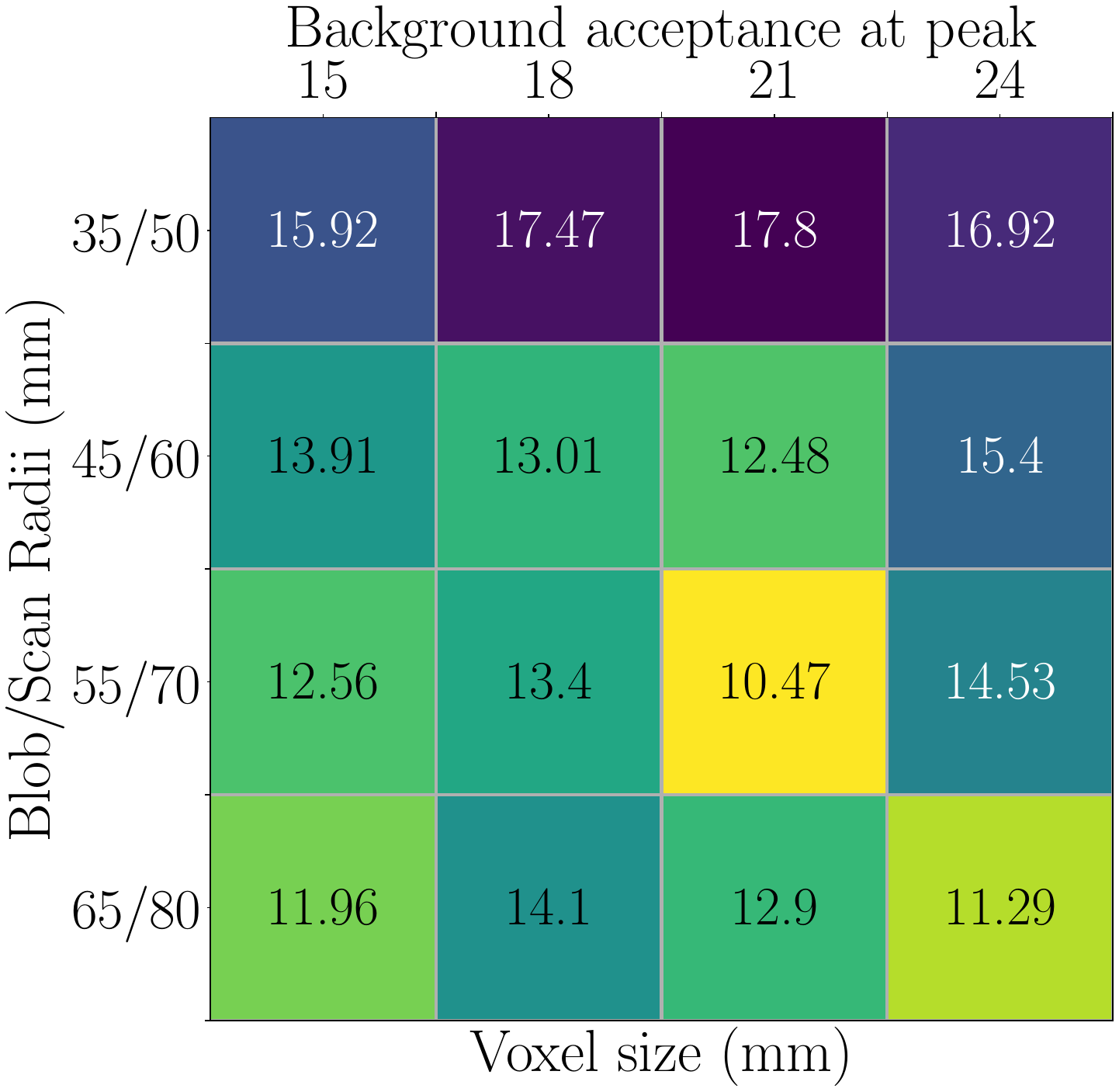}
    \end{subfigure}
    \caption{Parameter scan across voxel size and a pair of blob/scan radius parameters, with each quadrant illustrating differing performance metrics. Top left: maximal FOM Top right: Total number of events after selection cuts are applied as described in Section \ref{sec:DEP_ana}. Bottom left: signal efficiency at maximal FOM. Bottom right: background acceptance at maximal FOM.}
\end{figure}

\section{Changes in topological characteristics in data across differing pressures}\label{app:blob_radii}

This section compares topological characteristics observed in the double escape peak analysis of NEXT-100 (Section \ref{sec:DEP_ana}) with those from the equivalent analysis of NEXT-White \cite{topology_NEW}, which operated at $\sim$\qty{10}{bar}. The NEXT-100 LPR run operated at \qty{4}{bar}. This comparison is intended to be qualitative rather than quantitative, as the two detectors can not be considered like-for-like as discussed in Section~\ref{sec:discussion}. 

\subsection{Blob radius}\label{ssec:blob_rad}

\begin{figure}[ht]
    \includegraphics[width=\textwidth]{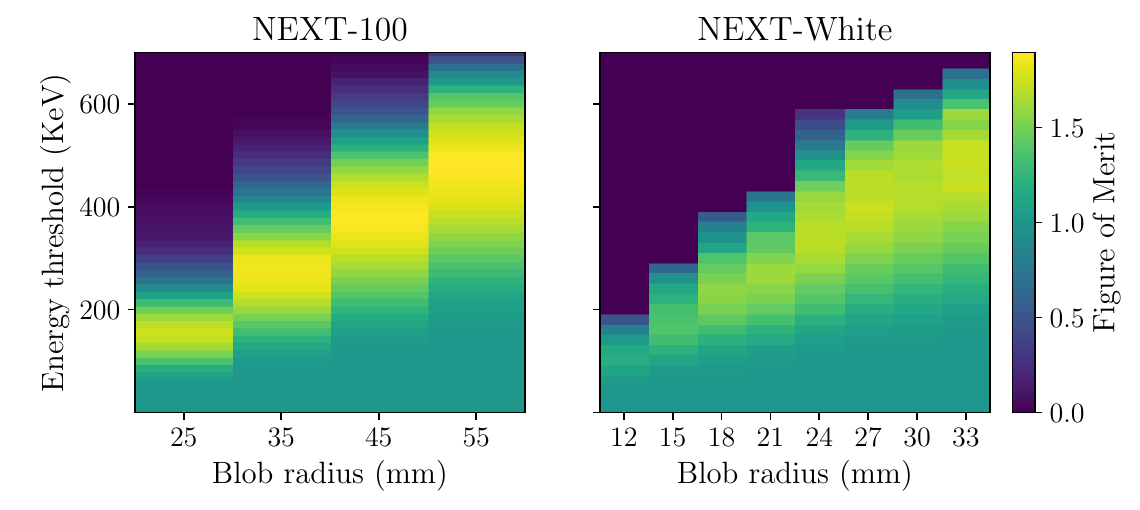}
    \caption{2D histogram of blob radius by energy threshold with the heatmap showing the figure of merit (FOM) values at each threshold value and blob radius. The data for NEXT-White at 10 bar (right) \cite{topology_NEW} is shown alongside data for NEXT-100 at 4 bar (left), demonstrating the larger blob radius required in the NEXT-100 LPR run to achieve comparable FOM distributions.}
    \label{fig:blobr_FOM_comparison}
\end{figure}

Figure \ref{fig:blobr_FOM_comparison} shows that a blob radius exceeding \qty{45}{mm} is required to achieve a maximal FOM with an energy threshold cut of approximately \qty{400}{keV} in NEXT-100, significantly larger than the \qty{25}{mm} required for NEXT-White, demonstrating that the blob radius requirement nearly doubles at lower pressure. It should be noted that the improved blob centring algorithm described in Section \ref{sec:data_and_topo} was applied to the NEXT-100 data but not to NEXT-White. If this improvement were removed, an even larger blob radius would be required to encapsulate the same energy, making the pressure-driven difference more pronounced.

\subsection{Track length}\label{ssec:track_length}

\begin{figure}[ht]
    \centering
    \begin{subfigure}{0.47\textwidth}
        \includegraphics[width=\textwidth]{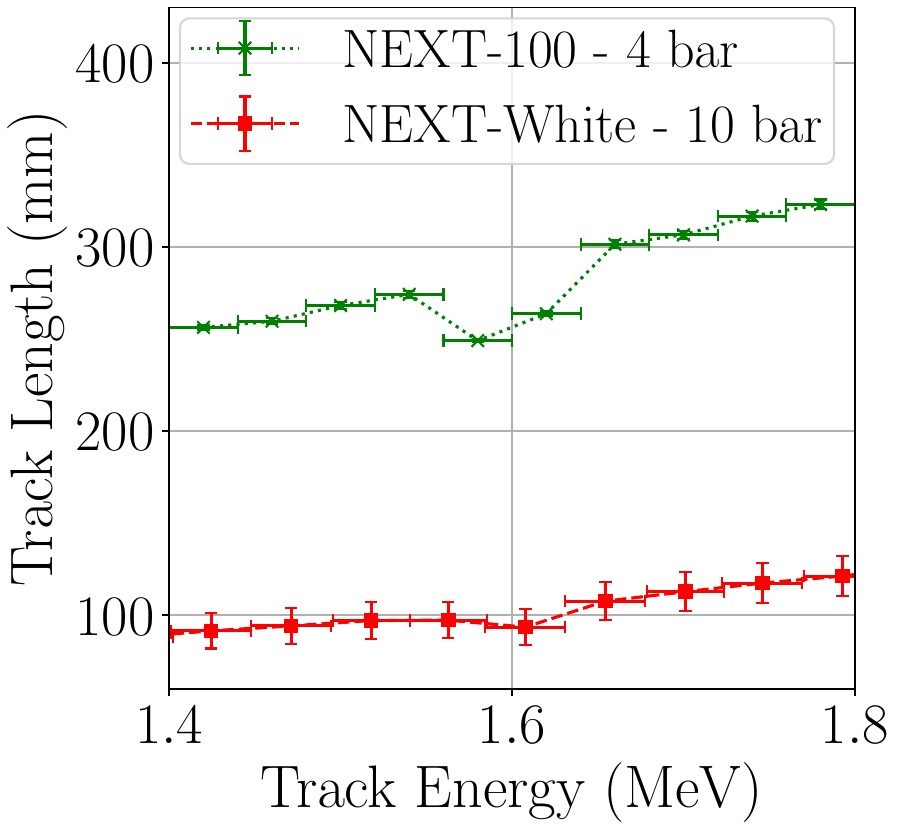}
    \end{subfigure}
    \hfill
    \begin{subfigure}{0.52\textwidth}
        \includegraphics[width=\textwidth]{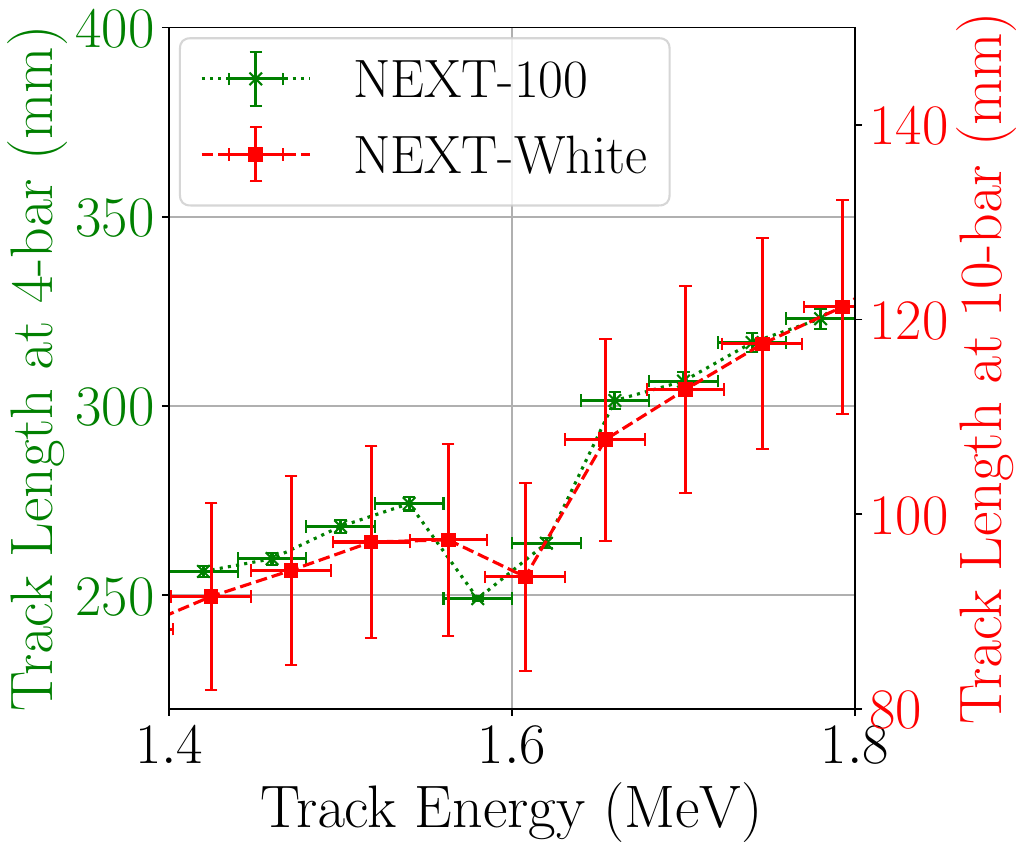}
    \end{subfigure}
    \caption{Track energy by track length for both the NEXT-White detector at \qty{10}{bar} and NEXT-100 at \qty{4}{bar}. Shown both on a shared track length axis (left), and independent axes (right) to demonstrate the shared distribution shape across track energy.}
    \label{fig:track_length_comparison}
\end{figure}

Similarly, a comparison of track length across the double escape peak energy range is shown in Figure \ref{fig:track_length_comparison}. A factor of 2-3 difference is observed across the two pressures, with the distinct bump at the double escape peak energy visible in both datasets.

\newpage
\section{Cut flow table}\label{app:cut_flow}
Provided here is a table of the efficiencies per cut applied across the analysis described in Section \ref{sec:DEP_ana}.

\begin{table}[ht]
\centering
\begin{tabular}{|l|l|}
\hline
\textbf{Cut Type}                                                     & \begin{tabular}[c]{@{}l@{}}Relative \\ Efficiency (\%)\end{tabular} \\ \hline
\textbf{Energy}                                                       & 11.87                                                               \\ \hline
\textbf{Fiducial}                                                     & 22.71                                                               \\ \hline
\textbf{One track}                                                    & 47.40                                                               \\ \hline
\textbf{\begin{tabular}[c]{@{}l@{}}Overlapping \\ blobs\end{tabular}} & 80.41                                                               \\ \hline
\end{tabular}
\caption{Table of cut efficiencies for cuts described in Section \ref{ssec:dep_ana_proper}. `Relative efficiency' in this context means that each efficiency is calculated with respect to the events before and after each cut is applied sequentially.} 
\label{tab:cut_flow}
\end{table}

The energy cut is applied prior to the topological analysis, as it can be applied prior to the topological reconstruction and significantly reduces runtime. 
The poor performance of the fiducial cut is primarily attributed to the calibration port being located at the radial maximum of the detector, resulting in the majority of events occurring near it. This, combined with the reduced pressure producing longer tracks that are more likely to exceed the fiducial volume, results in a poor efficiency.
The one track cut efficiency is decreased in comparison to prior works (66\% to 47\%) \cite{topology_NEW}, attributed to a lower specific energy loss. 
It should be noted however, that this efficiency is significantly improved relative to earlier stages of this analysis not discussed in detail here. 
These improvements were found primarily through two changes: the loosening of hit-based energy cuts which allowed for tracks with lower energy density (and resultantly less energetic hits) to be retained, and the implementation of improved-noise removal algorithms such as the clustering algorithm mentioned in Section \ref{ssec:light_detec_proc}. 
Together, these changes enable good one-track efficiency to be retained while avoiding both excessive track splitting and the loss of events due to noisy SiPMs producing fake multi-track events.
A significant increase in blob overlap is seen in comparison to the prior analysis applied for NEXT-White (2\% to 20\%) \cite{topology_NEW}. This is due to two factors: firstly, the new blob reconstruction algorithm described in Section \ref{sec:topoprocess} results in more overlapping blobs, as the new blob centres sit further from the track extrema (hence closer to one another) than previously. Secondly, due to the increased size of the blobs, a more conservative approach to reconstruction parameters was adopted to ensure that blob energies were fully encapsulated at the cost of some blob-overlap efficiency. 
In future analyses, it would be preferable to remove the blob-overlap cut entirely, and instead use an intelligent energy-distribution method to separate the energy of overlapping blobs. 

\bibliographystyle{JHEP}
\bibliography{biblio}

\end{document}